\documentstyle[11pt,aaspp4]{article}
\input psfig.tex
\def\aa{{ A\&A}}

\def\aj{{AJ}}
\def\al{$\alpha$}
\def\bet{$\beta$}

\def\apj{{ApJ}}
\def\apjs{{ApJS}}
\def\asec{$^{\prime\prime}$}

\def\deg{$^{\circ}$}

\def\e#1{ $\times$ 10$^{#1}$}
\def\etal{{et al. }}

\def\flamb{erg s$^{-1}$ cm$^{-2}$ \AA$^{-1}$}
\def\flux{ergs s$^{-1}$ cm$^{-2}$}

\def\hst{{\it HST}}
\def\kms    {~km~s$^{-1}$}
\def\lamb{$\lambda$}
\def\lum{ergs s$^{-1}$}

\def\mnras{{MNRAS}}
\def\nat{{Nature}}

\def\pasp{{PASP}}
\def\percm2{cm$^{-2}$}

\def\lax    {${_<\atop^{\sim}}$ }

\def\refindent{\par\noindent\parskip=2pt\hangindent=3pc\hangafter=1 }

\def\farcs{\hbox{$.\!\!^{\prime\prime}$}}

\def\oiii{[\ion{O}{3}]}

\def\oi{[\ion{O}{1}]}

\def\nii{[\ion{N}{2}]}

\def\sii{[\ion{S}{2}]}

\slugcomment{To appear in {\it The Astrophysical Journal}.}
\lefthead{Ho et al.}
\righthead{NGC 4450}

\begin{document}

\title{Double-Peaked Broad Emission Lines in NGC 4450 and Other 
LINERs\footnote{Based on observations with the {\it Hubble Space Telescope}, 
which is operated by AURA, Inc., under NASA contract NAS5-26555.}}

\author{Luis C. Ho\altaffilmark{2}, Greg Rudnick\altaffilmark{3}, Hans-Walter 
Rix\altaffilmark{4}, Joseph C. Shields\altaffilmark{5}, Daniel H. 
McIntosh\altaffilmark{3}, Alexei~V.~Filippenko\altaffilmark{6}, 
Wallace L. W. Sargent\altaffilmark{7}, and Michael Eracleous\altaffilmark{8}}

\altaffiltext{2}{The Observatories of the Carnegie Institution of Washington, 
813 Santa Barbara St., Pasadena, CA 91101-1292.}

\altaffiltext{3}{Steward Observatory, Univ. of Arizona, Tucson, AZ 85721.}

\altaffiltext{4}{Max-Planck-Institut f\"{u}r Astronomie, K\"{o}nigstuhl 17, 
Heidelberg, D-69117, Germany.}

\altaffiltext{5}{Ohio University, Dept. of Physics and Astronomy, Clippinger 
Labs 251B, Athens, OH 45701-2979.}

\altaffiltext{6}{Department of Astronomy, University of California, Berkeley, 
CA 94720-3411.}

\altaffiltext{7}{Palomar Observatory, 105-24 Caltech, Pasadena, CA 91125.}

\altaffiltext{8}{Department of Astronomy and Astrophysics, The
Pennsylvania State University, 525 Davey Lab, University Park, PA 16802.}

\begin{abstract}
Spectra taken with the {\it Hubble Space Telescope} (\hst) reveal that 
NGC 4450 emits Balmer emission lines with displaced double peaks and extremely 
high-velocity wings.  This characteristic line profile, previously seen in a 
few nearby LINERs and in a small fraction of broad-line radio galaxies, can be 
interpreted as a kinematic signature of a relativistic accretion disk.  We 
can reproduce the observed profile with a model for a disk with a radial 
range of 1000--2000 gravitational radii and inclined by 27 degrees
along the line of sight. The small-aperture \hst\ data also allow us to 
detect, for the first time, the featureless continuum at optical wavelengths 
in NGC 4450; the nonstellar nucleus is intrinsically very faint, with 
$M_B$ = --11.2 mag for $D$ = 16.8 Mpc.  

We have examined the multiwavelength properties of NGC 4450 collectively with 
those of other low-luminosity active nuclei which possess double-peaked broad 
lines and find a number of common features.  These objects are all classified 
spectroscopically as ``type 1'' LINERs or closely related objects.  The 
nuclear luminosities are low, both in absolute terms and relative to the 
Eddington rates.  All of them have compact radio cores, whose strength 
relative to the optical nuclear emission places them in the league of 
radio-loud active nuclei.  The broad-band spectral energy distributions of 
these sources are most notable for their deficit of ultraviolet emission 
compared to those observed in luminous Seyfert~1 nuclei and quasars.  The 
double-peaked broad-line radio galaxies Arp 102B and Pictor~A have very 
similar attributes.  We discuss how these characteristics can be understood 
in the context of advection-dominated accretion onto massive black holes.
\end{abstract}


\section{Introduction}

A minority of active galactic nuclei (AGNs) exhibit double-peaked broad 
emission lines --- permitted lines whose profile shows two displaced maxima, 
offset from the line center by several thousand \kms.  According to Eracleous 
\& Halpern (1994), approximately 10\% of broad-line radio galaxies show 
double-peaked lines.  A variety of mechanisms have been proposed to 
explain this unique kinematic signature, including relativistic motions in an 
accretion disk (Chen, Halpern, \& Filippenko 1989; Chen \& Halpern 1989), two 
separate broad-line regions due to a binary black hole (Gaskell 1983), 
biconical outflow (Zheng, Binette, \& Sulentic 1990), and anisotropic 
illumination of the broad-line region (Goad \& Wanders 1996).  
A combination of basic physical arguments and recent observational results, 
however, has shown that the most plausible origin of double-peaked emission
lines is the accretion disk around the central black hole. Alternative
interpretations are slowly being ruled out (Eracleous 1999).
 
A notable characteristic of double-peaked broad-line AGNs is that their 
distinctive line profile can be transient or highly variable.  In recent 
years, double-peaked broad emission lines have been found in several galaxies 
that previously had none.  These are NGC 1097 (Storchi-Bergmann, Baldwin, 
\& Wilson 1993), Pictor~A (Halpern \& Eracleous 1994; Sulentic et al. 
1995), and M81 (Bower et al. 1996).  That so many cases turn up 
serendipitously among nearby galaxies suggests that this is not an uncommon 
phenomenon.  An intriguing, possibly important, connection may exist with 
low-ionization nuclear emission-line regions (LINERs; Heckman 1980), a class 
of emission-line nuclei often found in nearby galaxies, but one whose nature 
remains controversial (Filippenko 1996; Ho 1999a).  The transient emission 
sources listed above all qualify spectroscopically as LINERs according to 
the low-ionization state of their optical spectra, as do the majority of the 
broad-line radio galaxies that contain double-peaked lines (e.g., Eracleous 
\& Halpern 1994).  
 
This paper reports the discovery of double-peaked broad emission lines 
in yet another LINER galaxy, NGC 4450, based on observations made with the 
{\it Hubble Space Telescope (HST)}.  NGC 4450 is a bright ($B_T$ = 10.9 mag) 
Sab galaxy located in the Virgo cluster.  The nucleus is spectroscopically 
classified by Ho, Filippenko, \& Sargent (1997a) as a LINER of ``type 1.9'' 
based on detection of weak broad H\al\ emission.

\section{Observations}

We observed NGC 4450 as part of a spectroscopic study of nearby spiral 
galaxies using the Space Telescope Imaging Spectrograph (STIS) on \hst\ 
(Rix et al. 2000).  Among the sample of 24 galaxies, NGC 4450 is one of two 
found to possess double-peaked broad lines; the other, NGC 4203, is discussed 
by Shields et al. (2000).  The observations of NGC 4450 were taken with the 
STIS CCD (Baum et al. 1999) on 1999 January 31 UT.  We used the 
0\farcs2$\times$52\asec\ slit, which for our program provides the optimal 
balance between signal-to-noise ratio and desired spatial resolution.  The 
angular scale of the chip is 0\farcs05 pixel$^{-1}$.  To  make scheduling more 
flexible, we did not preselect any specific orientation of the slit; at the 
time of the observations the position angle of the slit was at 53\deg, about 
58\deg\ from the major axis of the galaxy.  

The data presented here were obtained with two gratings.  The G750M grating 
covers $\sim$6300 to 6870 \AA\ at a full width at half-maximum (FWHM) spectral 
resolution of 0.84 \AA\ for a point source.  We also acquired 
spectra using the low-resolution G430L grating, which covers 3200--5700 \AA\ 
at a FWHM resolution of 4.1 \AA.  
Following a brief (30~s) target acquisition image using the F28X50LP 
long-pass filter, peak-up was performed directly on the galaxy nucleus, 
followed by three (900, 957, and 840~s) exposures with the G750M grating and 
two (840 and 829~s) exposures with the G430L grating.  Wavelength calibration 
frames were interleaved among the science exposures to monitor drifts in the 
dispersion scale.  Data reduction followed the steps outlined in Baum et al. 
(1999), which included subtraction of dark current, flat-fielding, removal of 
hot pixels and cosmic rays, correction for geometrical distortion, and flux 
and wavelength calibration.  A forthcoming paper (Rix et al. 2000) will 
describe in greater detail the observational strategy of our program and the 
procedures for data reduction.

\section{Results}

\subsection{Double-Peaked Broad Emission Lines}

Figure 1{\it a} shows the G750M spectrum of the nucleus of NGC 4450, extracted 
by adding the central three rows (corresponding roughly to the size of the 
point-spread function) of the summed two-dimensional frames.  The 
spectrum corresponds to an aperture of 0\farcs2$\times$0\farcs15.  Shown for 
comparison in Figure 1{\it b} is the spectrum of the object taken 15 years 
earlier (February 1984) by Filippenko \& Sargent (1985); note that the 
emission-line fluxes in the STIS aperture are much smaller than in the Palomar 
spectrum.  The G430L spectrum appears in Figure 1{\it c}.  
 
The most remarkable new feature in the STIS spectrum is the extremely broad, 
double-peaked or double-shouldered profile of the H\al\ line (Fig. 1{\it a}).
An elevated ``shelf'' of emission is apparent as well in H\bet, and possibly 
also in H$\gamma$ and H$\delta$ (Fig. 1{\it c}).  The feature was not evident 
in the earlier data of Filippenko \& Sargent (1985), nor was there any mention 
of it in the surveys of Stauffer (1982), Keel (1983), V\'eron-Cetty \& V\'eron 
(1986), or Rubin, Kenney, \& Young (1997).  These last authors observed 
NGC 4450 as recently as 1992 February.  Broad H$\alpha$ emission has previously
been noticed in NGC~4450 (Stauffer 1982; Filippenko \& Sargent 1985),
but at a level significantly weaker than found here.  These earlier
detections resemble ``normal'' broad-line emission characteristic of Seyfert~1
nuclei, as can be seen in Figure 1{\it b} (see also Fig. 12{\it f} of Ho et
al. 1997b), with a FWHM in this case of $\sim 2300$ km s$^{-1}$.  Our
simulations indicate that the double-shouldered component is sufficiently
prominent that it could have been detected in high-quality
ground-based observations, but would require a careful removal of the
underlying stellar continuum, which is often difficult in practice
(see Ho et al. 1997b).  For the Palomar data shown in Figure 1{\it b}, for
example, removal of wavelength-dependent focus variations resulted in
uncertainties in the continuum shape that could obscure the presence of
an emission component as broad as that seen in the {\it HST}
spectrum.  We conclude that the published data for NGC~4450 do not
provide a clear answer as to whether the double-shouldered broad emission
reported here is a transient or persistent feature.  The same remarks 
apply to the double-shouldered lines we found in NGC 4203 (Shields et al. 2000).

To better assess the broad, double-shouldered profile of H\al, we performed 
a multi-component decomposition of the G750M spectrum (Fig. 1{\it a}).  
After subtracting the continuum, which was reasonably well approximated by a 
constant, we generated a template profile from the \sii\ doublet in order to 
model \nii\ \lamb\lamb 6548, 6583 and the narrow H\al\ line. The 
\oi\ \lamb\lamb 6300, 6364 lines are significantly broader than \sii, and so 
they were fitted separately, each with a combination of two 
Gaussians.\footnote{The wide breadth of \oi\ relative to \sii\ is an indication
of density/velocity stratification in the narrow-line region, as first 
convincingly shown by Filippenko \& Halpern (1984) and Filippenko (1985). 
The fact that the \oi\ lines in the small-aperture STIS spectrum are 
broader than those in the Palomar spectrum indicates that the denser gas is 
indeed closer to the nucleus than the less dense gas, as expected.} The flux 
ratio and wavelength separation of the respective components of the doublets 
were constrained by atomic parameters.  The remaining broad (``normal'' plus 
double-shouldered) component of H\al\ can be 
represented by a combination of four Gaussians with FWHM $\approx$ 2000, 
3100, 4100, and 10,500 \kms, respectively.  The 2000 \kms\ component is 
centered near the peak of the narrow component of H\al; its width and 
strength are roughly comparable to those of the ``normal'' broad-line region 
component previously seen in ground-based data.  The two intermediate-width 
components correspond to the two displaced shoulders, while the extremely 
broad component covers the low-level, extended wings.  A similar decomposition 
was attempted for H\bet\ using \oiii\ \lamb 5007 as a model for the narrow 
lines.  The results in this case are more uncertain because of the lower 
signal-to-noise ratio and lower dispersion of the blue spectrum.  

The double-shouldered profiles of H\al\ and H\bet\ are plotted on a velocity 
scale in Figure 2.  The H\al\ line has an
enormous width: its FWHM is $\sim$9500 \kms, and the full width near
zero-intensity (FWZI) reaches at least 18,000 \kms.  Taking the peak of the
narrow component of H\al\ to be the systemic velocity of the nucleus, the
blue and red horns of the profile are offset by $-$1550 and $+$3500 \kms,
respectively, and the centroid of the broad profile is shifted by $+$320 \kms\
at FWHM and by $-$1600 \kms\ at FWZI.  The overall profile is clearly
asymmetric: the blue peak is 30\% higher than the red peak, and the red peak
has a more extended shoulder than the blue peak.  The integrated flux of the
line is 5.3\e{-14} \flux, or $L$(H\al) = 1.8\e{39} \lum\ for $D$ = 16.8 Mpc.
The profile of H\bet\ generally traces that of H\al, the only major difference
being that the two peaks of H\bet\ appear to have the same height.  The
integrated flux of H\bet\ is a factor of 4.8 weaker than H\al.

\begin{figure}
\hskip 0.5truein
\psfig{file=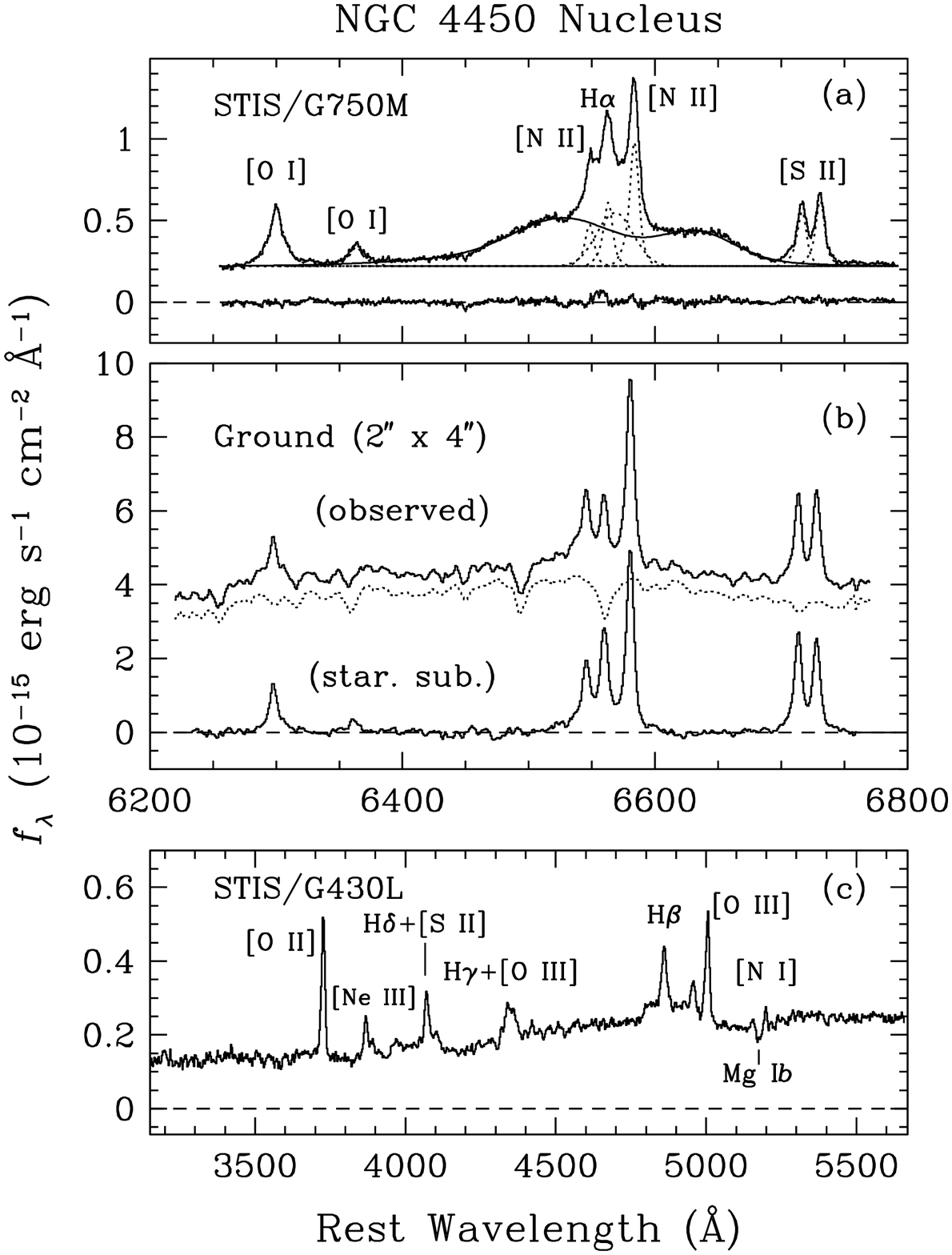,height=6.7truein}
\caption{
({\it a}) STIS G750M spectrum extracted from the central three rows 
(0\farcs2$\times$0\farcs15).  The dotted curves show profile fits for the 
narrow lines and for the ``normal'' broad component of H\al.  A single model 
profile was used for \nii, \sii, and narrow H\al, and another for 
\oi\ \lamb\lamb 6300, 6364.  The continuous curve, which consists of the 
sum of three Gaussians, is a simultaneous fit for the broad, double-shouldered 
component of H\al.  The residuals between the data and the fit are shown in 
the bottom plot.  ({\it b}) The Palomar 5~m spectrum (top), followed by
a template spectrum used to model the starlight (dotted line), and then the 
residual emission-line spectrum (bottom) obtained by subtracting the template
from the observed spectrum.  The observed and template spectra were offset
vertically by $-$2.8 and $-$3.3 units, respectively.  ({\it c}) STIS G430L 
spectrum extracted from the central three rows (0\farcs2$\times$0\farcs15).
}
\end{figure}
 
\begin{figure}
\plotone{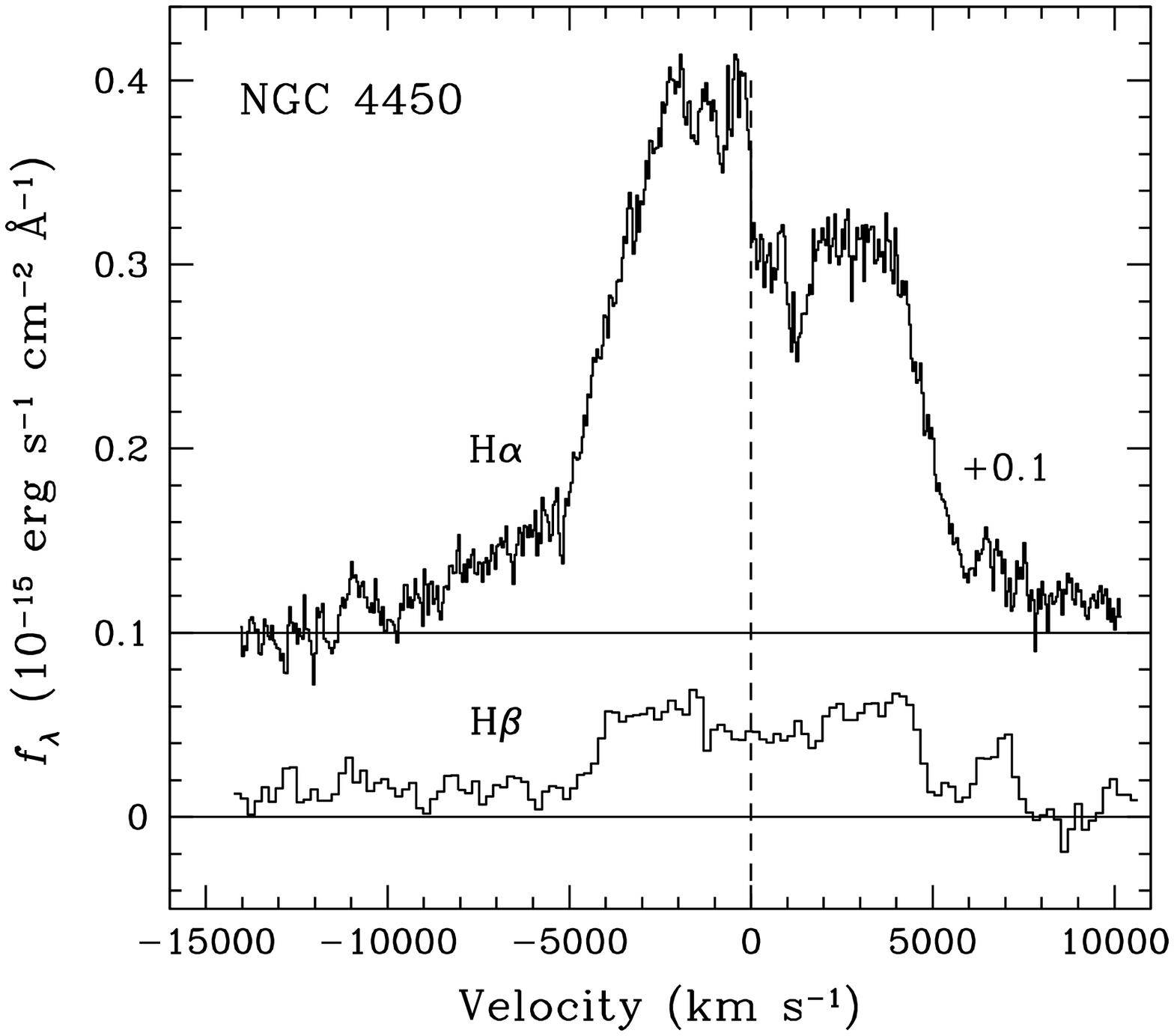}
\caption{
The broad, double-shouldered component of the H\al\ and H\bet\ lines plotted on
a velocity scale, whose origin is taken to be the peak of the narrow component
of H\al.  The narrow lines and the ``normal'' broad component of H\al\
have been removed (see Fig. 1{\it a}).  The narrow peak to the red of H\bet\
(at $\sim$7000 \kms) is due to imperfect subtraction of \oiii\
\lamb\lamb 4959, 5007.  The H\al\ curve has been shifted by +0.1 units in the
ordinate.
}
\end{figure}

\subsection{Nonstellar Featureless Continuum}

Whereas the ground-based spectrum of the nucleus is strongly contaminated
by the bulge light, requiring careful subtraction of the stellar background
in order to measure the emission lines accurately, the host galaxy has been
very effectively suppressed by the small aperture used in the \hst\
observations.  The equivalent widths of the emission lines in the
\hst\ spectrum are at least an order of magnitude larger than in the Palomar
spectrum, while the stellar absorption lines are much weaker. A significant
fraction of the continuum flux, therefore, originates from a featureless,
presumably nonstellar component associated with the active nucleus.  In
objects such as NGC 4450, which is quite representative of other LINERs
in nearby galaxies, the featureless continuum is virtually impossible to
detect from the ground.  We estimate the strength of the featureless continuum
from the dilution of the absorption features.  The strongest stellar line in
the region 6300--6900 \AA\ which is uncontaminated by nebular emission is
\ion{Ca}{1}+\ion{Fe}{1} \lamb6495, just blueward of \nii\ \lamb6548.  In the
nuclear regions of early-type spirals, this feature has an average equivalent
width of $\sim$0.9 \AA, and in the case of NGC 4450, its strength is 1.1 \AA\
(Ho et al. 1997a).  The line is not detected in the central row of the G750M
spectrum.  If we assume that the stellar population is the same as in 
the ground-based spectrum, we place an upper limit\footnote{We can 
rule out velocity broadening in the smaller STIS aperture as the cause for 
the apparent dilution of the \ion{Ca}{1}+\ion{Fe}{1} feature.  The 
best-fitting stellar template for the Palomar spectrum (Fig. 1{\it b}), 
measured with an aperture of 2\asec$\times$4\asec, has a central velocity 
dispersion of $\sigma_*$ = 130 \kms\ (Ho, Filippenko, \& Sargent, 
unpublished), consistent with values reported in the literature (McElroy 
1995).  We artificially broadened the template spectrum by different amounts to 
see the effects on the absorption lines.  Even for a hypothetical central 
velocity dispersion of $\sigma_*$ = 500 \kms, a very large value even 
if there is a massive black hole, the equivalent width of 
\ion{Ca}{1}+\ion{Fe}{1} diminishes only to 0.93 \AA.} of $\sim$0.4 \AA.
Therefore, at least 65\% of the continuum at 6500
\AA\ is nonstellar in the central 0\farcs2$\times$0\farcs05.  The G430L
spectrum shows a number of stellar absorption features, but again at
considerably reduced strengths compared to normal old stellar populations.
Judging from the equivalent widths of strong features such as the G~band
($\sim$4300 \AA) and \ion{Mg}{1}{\it b} ($\sim$5175 \AA), we estimate that
in the central 0\farcs2$\times$0\farcs05 approximately 75\% of the
continuum near 5000 \AA\ is contributed by a featureless component.  Assuming 
that this fraction is representative of the blue region of the spectrum, the 
nonstellar flux density at 4400 \AA\ is $f_\lambda$ = 7.7\e{-17} \flamb, which 
corresponds to $B$ = 19.9 mag, or $M_B$ = --11.2 mag for $D$ = 16.8 Mpc.

Independent evidence for a high nonstellar fraction in the nucleus of NGC 4450
comes from analysis of its central light distribution at optical and
ultraviolet (UV) wavelengths.  An archival F555W image of NGC 4450 taken with
the Second Wide-Field Planetary Camera (WFPC2) shows a distinct, point-like
nucleus with $m_{\rm F555W}\,\approx$ 19.0 mag (Ho et al. 2000b).  From the
conversion $m_{\rm F555W}$ = --2.5 log $f_\lambda$ $-$ 21.1 (Holtzman \etal
1995), $f_\lambda$ (5500 \AA) = 9.1\e{-17} \flamb, which is $\sim$40\% 
of the flux density measured in the central three rows of the G430L spectrum,
$f_\lambda$ (5500 \AA) = 2.4\e{-16} \flamb.  Taking into consideration the 
larger effective aperture of the STIS spectrum, this comparison indicates that 
a significant fraction of the light falling into the slit originates from an 
unresolved source, consistent with it being an AGN.  The nucleus has also been 
detected as a compact, point-like UV source in WFPC2 F218W (central wavelength 
$\sim$2200 \AA) images taken by Ho, Filippenko, \& Sargent (2000).
 
\section{Discussion}

\subsection{Evidence for an Accretion Disk in NGC 4450}

The double-shouldered H\al\ line in NGC 4450 strongly resembles the 
double-peaked profiles observed in some broad-line radio galaxies.   In those
objects the distinctive line profile has often been modeled in the context 
of a relativistic accretion disk (e.g., Chen et al. 1989; Eracleous \& Halpern 
1994; Halpern et al. 1996).  In this interpretation, the tremendous breadth of 
the line reflects the high rotation speed of the gas; the blue peak has 
greater intensity than the red peak because of relativistic beaming; and the 
overall asymmetry of the profile arises from the combined effects of 
transverse and gravitational redshift.  In NGC 4450, the observed profile of
H\bet\ differs slightly from that of H\al, but its signal-to-noise ratio is
considerably lower because of its reduced strength, and the difference between
H\al\ and H\bet\ is not inconsistent with that seen in broad-line radio
galaxies.

Figure 3 illustrates explicitly the viability of the disk interpretation for 
NGC 4450.  After subtracting the continuum, the H$\alpha$ profile was fitted 
with a model according to which the line originates in a relativistic
Keplerian disk (Chen et al. 1989; Chen \& Halpern
1989). In this model the line photons are emitted from a thin,
photoionized layer at the surface of a circular accretion disk, which
is illuminated by a source of ionizing radiation located at the center
of the disk. The calculation of the line profile includes, in addition
to Doppler broadening, the effects of transverse and gravitational
redshift and light bending. The free parameters of the model are the
inner and outer radii of the line-emitting portion of the disk,
$\xi_1$ and $\xi_2$ (expressed in units of the gravitational radius,
$r_{\rm g}\equiv GM/c^2$, where $M$ is the mass of the black hole),
the inclination angle of the disk axis to the line of sight, $i$, and
the broadening parameter, $\sigma$, which represents the velocity dispersion 
in a parcel of gas in the disk, presumably due to turbulence. The emissivity 
of the disk is assumed to vary with radius as $\epsilon\propto\xi^{-q}$, 
with $q$ set to 3, following the results of photoionization
calculations by Dumont \& Collin-Souffrin (1990a,b,c). The model
parameters that yield the best fit are $\xi_1=1010^{+610}_{-440}$,
$\xi_2=2030^{+1200}_{-990}$, $i=27^{+8}_{-7}$~degrees, and
$\sigma=1000^{+300}_{-200}~{\rm km~s}^{-1}$. These parameters are
comparable to those required to fit the profiles of double-peaked
emission lines found in broad-line radio galaxies (Eracleous \&
Halpern 1994) and in the LINER NGC 1097 (Eracleous et al. 1995;
Storchi-Bergmann et al. 1995, 1997).    

The bottom panel of Figure 3 shows the residual spectrum obtained by 
subtracting the disk model from the observed spectrum.  Note that the 
residual spectrum looks very similar to the starlight-subtracted 
ground-based spectrum shown in Figure 1{\it b}.

NGC 4203, the other double-peaked broad-line object discovered in our survey 
(Shields et al. 2000), is qualitatively very similar to NGC 4450 in its line 
profile.  Notably, the line profiles of both of these objects differ from 
those of NGC 1097 and M81, the two spirals that previously were found to 
emit transient double-peaked lines.  At the time of discovery by 
Storchi-Bergmann et al. (1993), the H\al\ line in NGC 1097 displayed a 
red peak stronger than the blue peak, and the blue shoulder was the more 
extended of the two.  The sense of the asymmetries, however, varied during 
subsequent monitoring, with the overall pattern interpretable in the 
context of a model of a precessing elliptical disk (Eracleous et al. 1995; 
Storchi-Bergmann et al. 1995, 1997).  When M81 was observed by Bower et al. 
(1996), it, too, had the initial appearance of NGC 1097.

\begin{figure}
\psfig{file=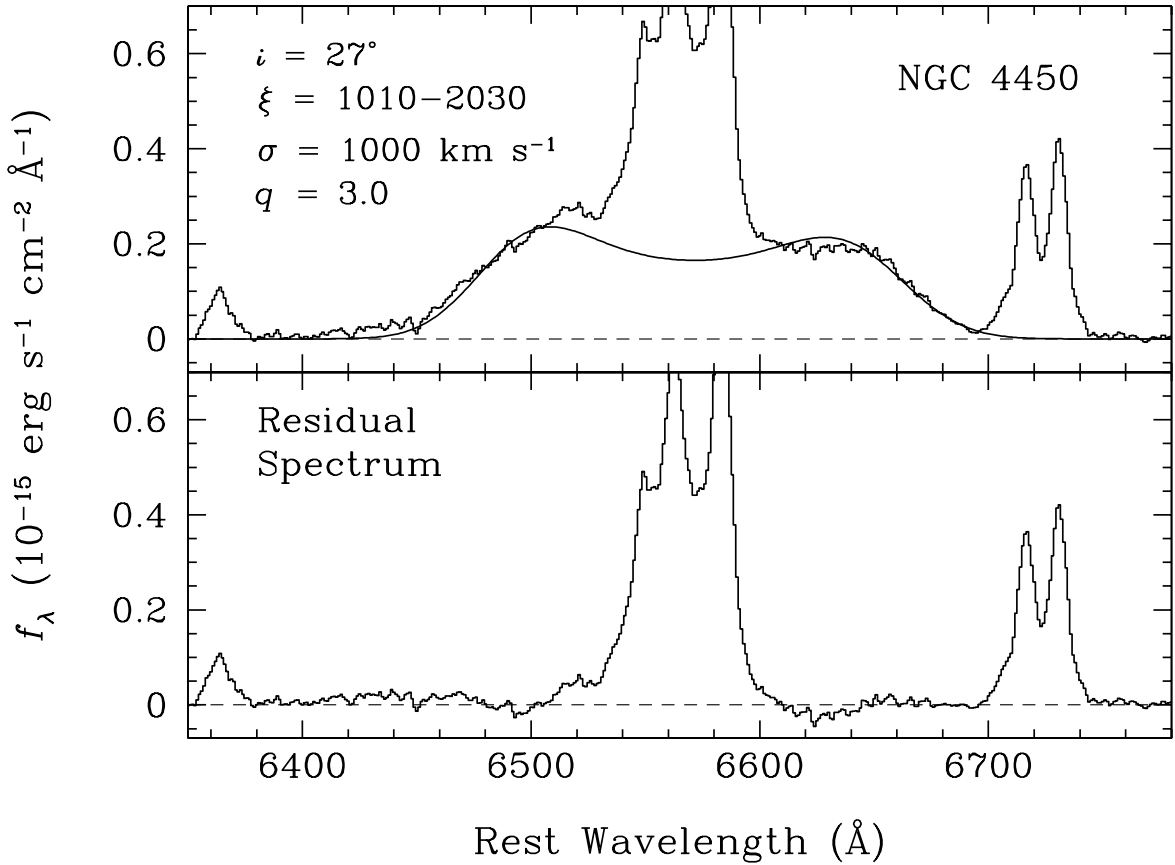,width=6.5truein,angle=270}
\caption{
The top panel shows an expanded view of the spectrum near the broad H\al\
line; the continuum has been removed.  The disk model described in the text is
overplotted as a continuous curve.  The bottom panel shows the residual
spectrum obtained by subtracting the disk model from the observed spectrum.
}
\end{figure}

\vbox{
\hbox{
\hskip -0.5truein
\psfig{file=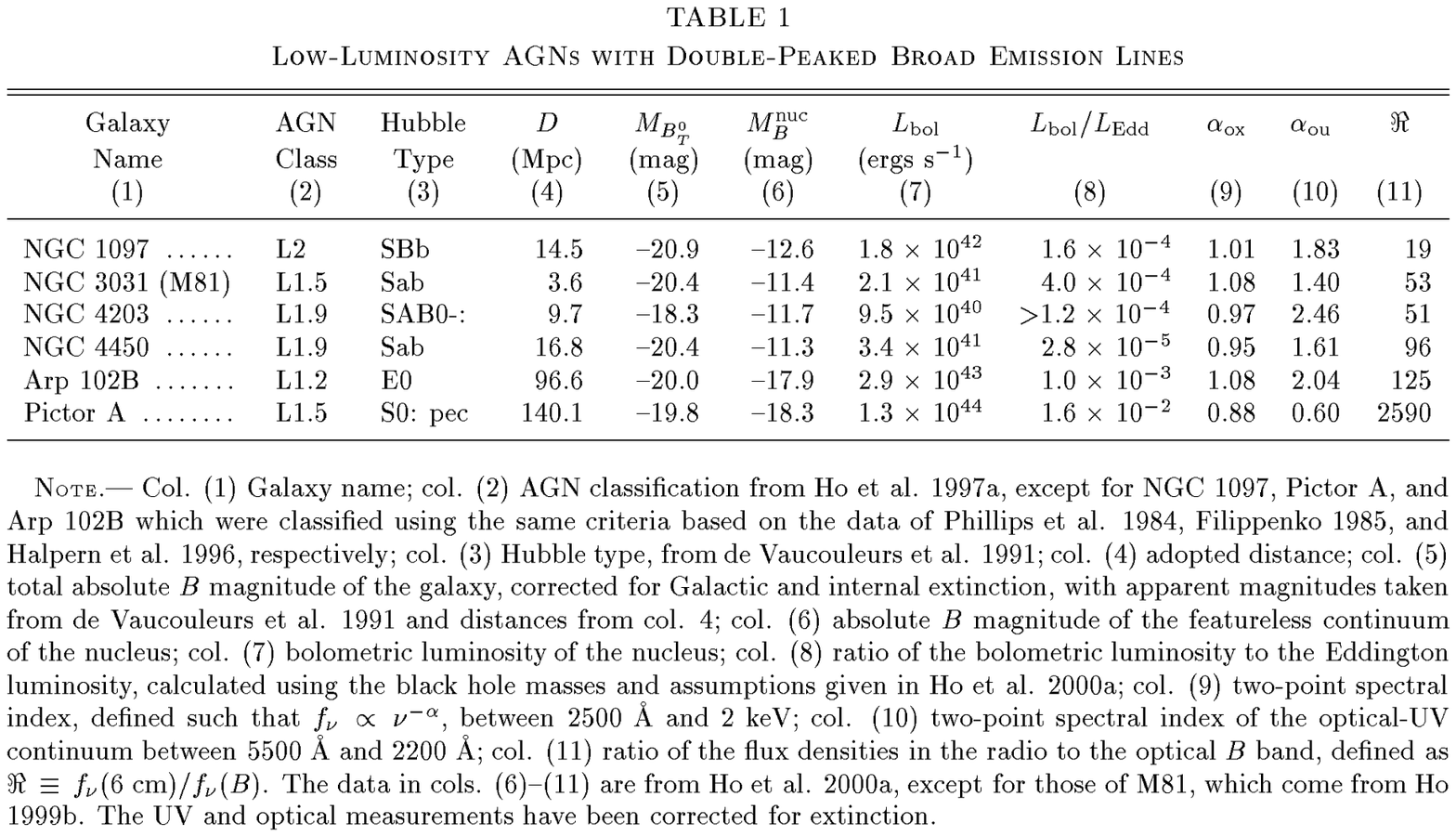,width=7.5truein}
}
}

\subsection{The Incidence and Detectability of Double-Peaked Broad Lines}

Double-peaked broad lines have recently been found serendipitously in
three cases, namely NGC 1097, M81, and Pictor~A (\S\ 1).  Our STIS
study adds two new members to the roster, NGC 4450 and NGC 4203.
Detection of such emission in low-luminosity sources is complicated by
its weak contrast with the underlying stellar continuum.  In the case
of the first three objects, detection was made possible by a transient
brightening of the broad feature.  For the last two sources, we do not
know whether variability contributed to their detection, but the use
of the STIS aperture probably did.  Indeed, the detection of two such
objects out of a sample of only 19 weak AGNs (Rix et al. 2000) is perhaps an 
indication of both the prevalence of such emission and the difficulty of 
discerning it from the ground, a point reinforced by the scarcity of such 
objects in ground-based surveys.

Disk-like line profiles are found preferentially in LINERs.  All the 
previously known transient cases, along with the two found in our survey,
are spectroscopically classified as LINERs (see Table 1), and the 
detection rate among LINERs in the STIS survey is 25\% (2/8).  The 
association with {\it type 1} LINERs --- those that have detectable emission 
from a broad-line region --- is especially striking.  With the exception of 
NGC 1097, all of the objects in Table 1 were known to be LINER~1s prior to 
the discovery of the double-peaked component, and two out of the four 
LINER~1s in the STIS sample turned out have disk-like profiles.  Consistent 
with this trend, Eracleous \& Halpern (1994) find that broad-line radio 
galaxies that show disk-like emission preferentially emit a stronger 
low-ionization, LINER-like spectrum. 

\subsection{Spectral Properties and Nature of the Accretion Flow}

Despite the limited number of known nearby low-luminosity AGNs with 
double-peaked broad emission lines, it is instructive to examine their 
spectral properties as a group.  Table 1 summarizes multiwavelength data 
available for NGC 1097, M81, NGC 4203, and NGC 4450, and, for comparison, 
the radio galaxies Pictor~A and Arp 102B.  The spectral energy distributions 
(SEDs) of these objects will be presented in greater detail by Ho et al. 
(2000a).  It is important to stress that the SED 
measurements pertain to the galaxy {\it nucleus}, which at virtually all 
wavelengths is much fainter than the host galaxy itself.  

We draw attention to a set of common spectral properties shared among the 
first four objects, all of which have nuclei with very low luminosities.  The 
absolute magnitudes of the nuclei range from $M_B^{\rm nuc}$  = --11.3 
to --12.6.  The most salient feature of the SEDs is their conspicuous deficit 
of UV emission, which manifests itself as continuum slopes in the optical--UV 
($\alpha_{\rm ou}$ = 1.4--2.5; see Table 1 for specific definitions) and in 
the UV--X-ray ($\alpha_{\rm ox}$ = 0.9--1.1) regions that are steeper and 
flatter, respectively, than is found in luminous AGNs ($\alpha_{\rm ou}\, 
\approx$ 0.2--0.5, Malkan 1988; $\alpha_{\rm ox}\,\approx$ 1.2--1.6, 
Mushotzky \& Wandel 1989).  Following the arguments of Ho (1999b), we believe 
that the UV deficit is intrinsic to the source and not an artifact of dust 
extinction.  This suggests that the ``big blue bump'' (Shields 1978; Malkan 
\& Sargent 1982), and by implication the optically thick, geometrically thin 
accretion disk thought to give rise to this feature, is weak or absent.  The 
strength of the radio band is also notable.  All four objects, among them 
three spirals, contain compact, flat-spectrum radio cores.  The absolute radio 
power of the nuclei is tiny by traditional standards, with $P_{\rm 6~cm}\, 
\approx\,10^{20}$ W Hz$^{-1}$, but their optical output is correspondingly 
low, such that the normalized radio power remains relatively high.  The ratio 
$\Re$, defined as $f_{\nu}({\rm 6~cm})/f_{\nu}(B)$, spans $\sim$20--100, to 
be compared with $\Re\,>$ 10 for radio-loud quasars (Kellermann \etal 1989; 
Visnovsky \etal 1992).  By this conventional criterion, therefore, all four 
sources formally qualify as radio-loud, a finding that refutes the popular 
prejudice that radio-loud AGNs are confined to luminous early-type hosts.  We 
note, however, that none of these low-luminosity objects possess 
well-collimated, large-scale radio jets. 

The accretion flow in the systems highlighted here may be systematically 
different from the standard Shakura \& Sunyaev (1973) disk normally assumed to 
exist in luminous AGNs.  Recently there has been increasing theoretical 
and observational evidence that black holes fed at a low rate accrete through 
an optically thin, advective mode (see review by Narayan, Mahadevan, \& 
Quataert 1998).  ``Advection-dominated'' accretion flows (ADAFs) have several 
characteristics that make them an attractive framework in which to interpret 
our observations.  First, ADAFs are typified by a low radiative efficiency 
($L\,\propto\,\dot{m}^2$ instead of $\dot{m}$), and so they provide a 
natural explanation for the very low accretion luminosities actually 
measured.  The nuclei of the four nearby sources in Table 1, for instance, have
bolometric luminosities of $L_{\rm bol}\,\approx\, 10^{41}-10^{42}$ \lum.  
Second, the ADAF phase can only exist in sub-Eddington systems, when 
$L_{\rm bol}/L_{\rm Edd}$ \lax $10^{-2}$--10$^{-1}$ (Narayan et al. 1998); 
this condition is fulfilled by our sources, which have 
$L_{\rm bol}/L_{\rm Edd},\approx$ 1\e{-4} -- 3\e{-5} (Table 1).  Third, ADAFs 
do not emit a thermal UV bump, in qualitative agreement with the observed 
SEDs.  Fourth, thermal synchrotron radio emission contributes significantly 
to the ADAF spectrum, and consequently we expect the SED to be prominent in the 
radio.  Finally, we note that the characteristically harder ionizing spectrum 
of an ADAF lowers the effective ionization parameter and hence favors the 
production of a LINER-like emission-line spectrum (Halpern \& Steiner 1983; 
Ferland \& Netzer 1983).


The picture we are advocating for low-luminosity AGNs with double-peaked broad 
lines is likely to be more widely applicable to low-luminosity AGNs as a class.
The characteristics of the SEDs discussed above are not restricted to the 
specific sample emphasized in this study; they were noted by Ho (1999b) in 
a number of other low-luminosity systems, among them M81.  Quataert et al. 
(1999) have successfully applied ADAF models to fit the SEDs of M81 and 
NGC 4579, the latter a LINER not known to possess disk-like broad lines but 
otherwise very similar to M81.  

The basic elements of the above scenario may also be appropriate in general 
for radio galaxies with disk-like broad emission lines.  We illustrate this 
connection in Table 1 by comparing the first four nearby, low-luminosity 
systems with two well-studied, double-peaked broad-line radio galaxies, 
Arp 102B and Pictor~A.  Although both objects are substantially more 
luminous than the rest ($M_B^{\rm nuc}$ = --17.9 mag for Arp 102B; 
$M_B^{\rm nuc}$ = --18.3 mag for Pictor~A), they still rank among the least 
luminous classical Seyferts known.  More germane to the present discussion is 
the Eddington ratio, which is $\sim$1\e{-3} for Arp 102B and 2\e{-2} for 
Pictor~A, both within the threshold in which the ADAF framework holds.
In their study of Arp 102B, Chen et al. (1989) already suggested that 
the structure of its accretion flow may take the form of an ``ion-supported 
torus'' (Rees et al. 1982), the physical concept of which is very similar to 
that of an ADAF.  Eracleous \& Halpern (1994) subsequently extended this idea 
to interpret a number of statistical properties observed in disk-like 
emitters.  However, one of the key pieces of evidence --- the absence of the 
UV bump --- has, until very recently, remained elusive.  Halpern \etal (1996) 
found that the \hst\ spectrum of Arp 102B indeed shows weak UV emission.  As 
shown in Table 1, its continuum has a steep optical--UV slope 
($\alpha_{\rm ou}$ = 2.04) and a relatively hard UV to X-ray slope 
($\alpha_{\rm ox}$ = 1.08).  Pictor~A, too, has an 
exceptionally flat UV--X-ray slope ($\alpha_{\rm ox}$ = 0.88), although 
its optical--UV slope ($\alpha_{\rm ou}$ = 0.60) does not differ 
markedly from those typically seen in luminous AGNs.  This may reflect 
the fact that its Eddington ratio ($\sim$2\e{-2}), the largest in the 
sample, is only marginally within the ADAF limit.

\section{Summary}

Double-peaked, broad emission lines in AGNs, often interpreted as a 
kinematic signature of a relativistic accretion disk, may be much more 
common than previously thought.  An \hst\ spectroscopic survey of a small 
sample of nearby galaxies discovered two such objects.  Here we present the 
case of NGC 4450; the other, NGC 4203, is discussed by Shields et al. 
(2000).  We show that the broad, double-peaked profile of the H\al\ line 
in NGC 4450 can be successfully fitted with a model for an inclined disk with a 
radial extent of 1000--2000 gravitational radii. 

We have collected small-aperture multiwavelength data for the few nearby 
low-luminosity AGNs known to possess double-peaked emission lines, and 
also for two higher luminosity counterparts, Arp 102B and Pictor~A,
in order to investigate their SEDs.  This study, in conjunction 
with that of Ho (1999b), substantially increases the number of faint nuclei 
having broad-band nuclear SEDs.  We argue that these objects share a set of 
common spectral characteristics which can be understood in the context of 
advection-dominated accretion onto massive black holes.

\acknowledgments
This work was partially supported by NASA grants NAG 5-3556 and GO-07361, the 
latter awarded by the Space Telescope Science Institute, which is operated by 
AURA, Inc., under NASA contract NAS5-26555.  We thank C.~Y. Peng for his help 
in fitting the line profiles and in analyzing the \hst\ images of NGC 4203 and 
NGC 4450, L. Colina for supplying the UV data of NGC 1097 in advance of 
publication, and V.~C. Rubin for sharing her ground-based spectra of NGC 4450. 



\vskip 0.70truein
\centerline{\bf{References}}
\medskip

\refindent 
Baum, S.~A., \etal 1999, STIS Instrument Handbook, Version 3 (Baltimore: STScI)

\refindent 
Bower, G.~A., Wilson, A.~S., Heckman, T.~M., \& Richstone, D.~O. 1996, \aj,
111, 1901

\refindent 
Chen, K., \& Halpern, J.~P. 1989, \apj, 344, 115

\refindent 
Chen, K., Halpern, J.~P., \& Filippenko, A.~V. 1989, \apj, 339, 742

\refindent 
de Vaucouleurs, G., de Vaucouleurs, A., Corwin, H.~G., Jr., Buta, R.~J.,
Paturel, G., \& Fouqu\'e, R. 1991, Third Reference Catalogue of Bright
Galaxies (New York: Springer)

\refindent 
Dumont, A. M., \& Collin-Souffrin, S. 1990a, A\&A, 229, 302

\refindent 
Dumont, A. M., \& Collin-Souffrin, S.  1990b, A\&A, 229, 313

\refindent 
Dumont, A. M., \& Collin-Souffrin, S.  1990c, A\&AS, 83, 71

\refindent 
Eracleous, M. 1999, in Structure and Kinematics of Quasar Broad-Line Regions,
ed. C.~M. Gaskell et al. (San Francisco: ASP), 163

\refindent 
Eracleous, M., \& Halpern, J.~P. 1994, \apjs, 90, 1

\refindent 
Eracleous, M., Livio, M., Halpern, J.~P., \& Storchi-Bergmann, T. 1995, \apj,
438, 610

\refindent 
Ferland, G. J., \& Netzer, H. 1983, \apj, 264, 105

\refindent 
Filippenko, A.~V. 1985, \apj, 289, 475

\refindent 
Filippenko, A.~V. 1996, in The Physics of LINERs in View of Recent 
Observations, ed. M. Eracleous et al. (San Francisco: ASP), 17

\refindent 
Filippenko, A.~V., \& Halpern, J.~P. 1984, \apj, 285, 458

\refindent 
Filippenko, A.~V., \& Sargent, W.~L.~W. 1985, \apjs, 57, 503

\refindent 
Gaskell, C.~M. 1983, in Quasars and Gravitational Lenses, Proc. 24th Li\'ege 
Astrophysical Colloquium (Li\'ege: Institute d'Astrophysique, Univ. Li\'ege), 
473

\refindent 
Goad, M.~R., \& Wanders, I. 1996, \apj, 469, 113

\refindent 
Halpern, J.~P., \& Eracleous, M. 1994, \apj, 433, L17

\refindent 
Halpern, J.~P., Eracleous, M., Filippenko, A.~V., \& Chen, K. 1996, \apj, 464,
704

\refindent 
Halpern, J.~P., \& Steiner, J.~E. 1983, \apj, 269, L37

\refindent 
Ho, L.~C. 1999a, Advances in Space Research, 23 (5-6), 813

\refindent 
Ho, L.~C. 1999b,  \apj, 516, 672

\refindent 
Ho, L.~C., \etal 2000a, in preparation

\refindent 
Ho, L.~C., Filippenko, A.~V., \& Sargent, W.~L.~W. 1995, \apjs, 98, 477

\refindent 
Ho, L.~C., Filippenko, A.~V., \& Sargent, W.~L.~W. 1997a, \apjs, 112, 315

\refindent 
Ho, L.~C., Filippenko, A.~V., \& Sargent, W.~L.~W. 2000, in preparation

\refindent 
Ho, L.~C., Filippenko, A.~V., Sargent, W.~L.~W., \& Peng, C.~Y. 1997b, \apjs,
112, 391

\refindent
Ho, L.~C., Peng, C.~Y., Filippenko, A.~V., \& Sargent, W.~L.~W. 2000b, in
preparation

\refindent
Holtzman, J., \etal 1995, \pasp, 107, 1065

\refindent 
Kellermann, K.~I., Sramek, R.~A., Schmidt, M., Shaffer, D.~B., \& Green,
R.~F. 1989, \aj, 98, 1195

\refindent 
Malkan, M.~A. 1988, Advances in Space Research, 8 (2-3), 49

\refindent 
Malkan, M.~A., \& Sargent, W.~L.~W. 1982, \apj, 254, 22

\refindent
McElroy, D.~B. 1995, \apjs, 200, 105

\refindent
Mushotzky, R.~F., \& Wandel, A. 1989, \apj, 339, 674

\refindent 
Narayan, R., Mahadevan, R., \& Quataert, E. 1998, in The Theory of Black Hole
Accretion Discs, ed.  M. A. Abramowicz, G. Bj\"{o}rnsson, \& J. E. Pringle
(Cambridge: Cambridge Univ. Press), 148

\refindent 
Phillips, M.~M., Pagel, B.~E.~J., Edmunds, M.~G., \& D\'\i az, A. 1984, \mnras,
210, 701

\refindent 
Quataert, E., Di Matteo, T., Narayan, R., \& Ho, L.~C. 1999, \apj, 525, L89

\refindent 
Rees, M.~J., Begelman, M.~C., Blandford, R.~D., \& Phinney, E.~S. 1982,
\nat, 295, 17

\refindent 
Rix, H.-W., \etal 2000, in preparation

\refindent 
Rubin, V.~C., Kenney, J.~D.~P., \& Young, J.~S. 1997, \aj, 113, 1250

\refindent 
Shakura, N.~I., \& Sunyaev, R.~A. 1973, \aa, 24, 337

\refindent 
Shields, G.~A. 1978, \nat, 272, 706

\refindent 
Shields, J.~C., Rix, H.-W., McIntosh, D.~H., Ho, L.~C., Rudnick, G., 
Filippenko, A.~V., Sargent, W.~L.~W., \& Sarzi, M. 2000, \apj, in press 

\refindent 
Stauffer, J.~R. 1982, \apj, 262, 66

\refindent 
Storchi-Bergmann, T., Baldwin, J.~A., \& Wilson, A.~S. 1993, \apj, 410, L11

\refindent 
Storchi-Bergmann, T., Eracleous, M., Livio, M., Wilson, A.~S.,
Filippenko, A.~V., \& Halpern, J.~P. 1995, \apj, 443, 617
 
\refindent 
Storchi-Bergmann, T., Eracleous, M., Ruiz, M.~T., Livio, M., Wilson, 
A.~S., \& Filippenko, A.~V. 1997, \apj, 489, 87

\refindent 
Sulentic, J.~W., Marziani, P., Zwitter, T., \& Calvani, M. 1995, \apj, 438, L1

\refindent 
Sulentic, J.~W., Zheng, W., Calvani, M., \& Marziani, P. 1990, \apj, 355, L15

\refindent 
Visnovsky, K.~L., Impey, C.~D., Foltz, C.~B., Hewett, P.~C., Weymann,
R.~J., \& Morris, S.~L. 1992, \apj, 391, 560

\refindent 
Zheng, W., Binette, L., \& Sulentic, J.~W. 1990, \apj, 365, 115

\end{document}